\let\kernelLabel\label
\documentclass[10pt,aps,prd,nofootinbib,superscriptaddress,twocolumn,floatfix,showkeys]{revtex4-2}
\let\label\kernelLabel
\usepackage[utf8]{inputenc}
\usepackage[T1]{fontenc}
\usepackage{microtype}
\usepackage{amsmath,amssymb,bm,mathtools}
\usepackage{graphicx}
\usepackage{booktabs,multirow,tabularx}
\usepackage{xcolor}
\usepackage[colorlinks=true,citecolor=blue,linkcolor=blue,urlcolor=blue]{hyperref}
\usepackage[normalem]{ulem}
\usepackage[capitalize,nameinlink]{cleveref}
\newcommand{\aone}{\alpha_1}
\newcommand{\atwo}{\alpha_2}
\newcommand{\athree}{\alpha_3}

\begin{document}
\title{Testing Extra-Dimensional Gravitational-Wave Polarizations with Compact-Binary Observations}
\author{Hai-Tian Wang}\email{wanght9@dlut.edu.cn}
\affiliation{School of Physics, Dalian University of Technology, Dalian 116024, China}
\author{Dicong Liang}\email{dcliang@smu.edu.cn}
\affiliation{Department of Mathematics and Physics, School of Biomedical Engineering, Southern Medical University, Guangzhou 510515, China}
\author{Lijing Shao}\email{lshao@pku.edu.cn}
\affiliation{Kavli Institute for Astronomy and Astrophysics, Peking University, Beijing 100871, China}
\affiliation{National Astronomical Observatories, Chinese Academy of Sciences, Beijing 100101, China}
\author{Yu-Xiao Liu}\email{liuyx@lzu.edu.cn}
\affiliation{Lanzhou Center for Theoretical Physics, Key Laboratory of Theoretical Physics of Gansu Province, and Key Laboratory of Quantum Theory and Applications of MoE, Lanzhou University, Lanzhou 730000, China}
\affiliation{Institute of Theoretical Physics \& Research Center of Gravitation, Lanzhou University, Lanzhou 730000, China}

\begin{abstract}

Extra spatial dimensions arise naturally in theories beyond four-dimensional general relativity (GR), but direct observational evidence remains elusive. Gravitational waves offer a distinctive probe: geometric projection onto the observable three-dimensional spatial subspace can produce apparent tensor, vector, and scalar polarizations. We consider one additional spatial dimension and a two-mode tensor-wave ansatz for which the projected polarization components are fixed linear combinations of the same two underlying GR tensor waveforms. These components enter the measured strain through the standard detector antenna patterns. A frequency-domain Bayesian analysis of five compact-binary events constrains two rotation angles for two noncommuting orderings of rotations that preserve propagation within the observed subspace. Two events are further analyzed for four orderings that allow propagation to tilt into the extra dimension. For GW250114, the maximum GR quantiles among the tested rotation orderings are $\mathcal Q_{\mathrm{GR}}\simeq0.6\%$ for the two-angle fits and $\mathcal Q^{\mathrm{marg}}_{\mathrm{GR}}\simeq4.6\%$ for the three-angle fits after marginalizing over the third angle. Across the full event sample, the GR reference point lies inside the estimated 90\% highest-posterior-density region in every case.


\end{abstract}
\maketitle

\section{Introduction}\label{sec:introduction}
Extra spatial dimensions offer a route to unifying gravity with the other fundamental interactions. They arise in Kaluza--Klein theory and string theory \cite{Kaluza:1921tu,Klein:1926tv,Overduin:1997sri,Gross:1984dd,Candelas:1985en,Cvetic:2022fnv}, and underpin phenomenological scenarios including large, warped, universal, and infinite-volume extra dimensions \cite{Arkani-Hamed:1998jmv,Randall:1999ee,Randall:1999vf,Appelquist:2000nn,Hooper:2007qk,Dvali:2000hr}. Their signatures have been sought through inverse-square-law tests \cite{Hoyle:2000cv,Kapner:2006si,Kehagias:1999my,Floratos:1999bv,Callin:2005wi} and collider searches for missing energy, virtual graviton exchange, and Kaluza--Klein resonances \cite{Giudice:1998ck,Mirabelli:1998rt,Han:1998sg,Hewett:1998sn,Davoudiasl:1999jd}. Related collider phenomenology includes pair production of Kaluza--Klein states, cascade decays, and additional graviton channels \cite{Cheng:2002ab,Datta:2005zs,Agashe:2007zd}, while monojet, diphoton, and missing-energy observations have constrained the respective scenarios \cite{ATLAS:2011kno,CMS:2012ucb,CMS:2011uvc,CMS:2011bsw,Deutschmann:2017bth}.

Astrophysical and cosmological observations probe complementary regimes. Examples include energy loss through Kaluza--Klein gravitons in SN1987A \cite{Cullen:1999hc,Hanhart:2000er}, mixing with bulk sterile neutrinos \cite{Caldwell:2000zn,Davoudiasl:2002fq}, relic production and modified cosmic expansion \cite{Hall:1999mk,Fairbairn:2001ct,Lue:2005ya,Li:2005aia}, a possible contribution of graviton emission to the knee in the cosmic-ray spectrum \cite{Kazanas:2001ep}, and strong-lensing or black-hole-shadow signatures \cite{Chakraborty:2016lxo,Lemos:2024wwi}. Gravitational waves (GWs) extend these tests to the dynamical strong-field regime. Proposed signatures include cosmic-superstring backgrounds, Kaluza--Klein spectral lines, modified phase evolution, and shifts of black-hole quasinormal modes \cite{Damour:2004kw,Seahra:2004fg,Andriot:2017oaz,Du:2020rlx,Mishra:2021waw}. Multimessenger and standard-siren observations have also constrained bulk shortcuts, amplitude leakage, and effective spacetime dimensionality \cite{Visinelli:2017bny,Pardo:2018ipy,LIGOScientific:2018dkp,Lin:2023zna,Corman:2021avn}.

Here we test a qualitatively different signature: apparent extra polarization content generated solely by geometric projection. 
A higher-dimensional tensor wave can project onto the observed three-dimensional spatial subspace as tensor, vector, breathing, and longitudinal components \cite{Liu:2022cwb}. 
We model the two natural-frame tensor amplitudes with standard four-dimensional GR compact-binary waveforms. Their linear combinations determine the evolution of all projected components. 
This correlated structure differs from hypotheses that introduce independently excited scalar or vector degrees of freedom in four-dimensional modified gravity \cite{Eardley:1973zuo,Liang:2017ahj,Hou:2017bqj,Moretti:2019yhs,Gong:2018cgj,Liang:2022hxd,Dong:2023bgt,Dong:2023xyb,Dong:2024zal}.

Existing polarization tests have found no significant evidence for generic non-tensorial modes \cite{LIGOScientific:2019fpa,Hagihara:2019ihn,LIGOScientific:2020tif,LIGOScientific:2021sio,LIGOScientific:2026qni,LIGOScientific:2026oim}. 
Higher-dimensional projection specifies a correlated polarization structure for these tests. We therefore construct the projected polarization waveforms, combine them using the interferometer antenna patterns, and infer the orientation angles jointly with the binary parameters, including spins. 
We analyze five compact-binary events under two rotation orderings that preserve propagation in the observed subspace, and use two events to test four additional orderings that tilt the propagation direction into the extra dimension. We find no evidence for projection-induced extra polarizations in the analyzed events.

\section{Projection-induced polarizations}\label{sec:model}
We consider a five-dimensional spacetime with one additional spatial coordinate, $w$. 
We assume that the underlying GW waveform is described by the two tensor polarizations of four-dimensional GR, while the additional spatial dimension modifies how these modes are projected onto the observer's three-dimensional subspace.
In the natural wave frame, the wave propagates along $z$, and its two tensor polarizations are confined to the transverse $x-y$ plane. Thus, its spatial polarization tensor is
\begin{equation}
p=\begin{pmatrix}h_+&h_\times&0&0\\h_\times&-h_+&0&0\\0&0&0&0\\0&0&0&0\end{pmatrix},
\label{eq:natural_tensor}
\end{equation}
where the coordinates are $(x,y,z,w)$. 
We parametrize the relative orientation of this frame and the observer's spatial hypersurface by rotations $R_{xw}(\aone)$, $R_{yw}(\atwo)$, and $R_{zw}(\athree)$. 
The first two tilt the transverse polarization plane into the extra dimension, while the third tilts the propagation direction. 
The rotated tensor is $\widetilde p=R p R^{\mathsf T}$. Its restriction to the observed subspace is
\begin{equation}
{}^{(3)}\widetilde p=\begin{pmatrix}
\widetilde h_+ +\widetilde h_b&\widetilde h_\times&\widetilde h_x\\
\widetilde h_\times&\widetilde h_b-\widetilde h_+&\widetilde h_y\\
\widetilde h_x&\widetilde h_y&\widetilde h_l
\end{pmatrix}.
\label{eq:projected_tensor}
\end{equation}
Here, $b$ and $l$ label the breathing and longitudinal scalar modes, respectively, while $x$ and $y$ label the two vector modes.
Each projected amplitude is a deterministic linear combination
\begin{equation}
\widetilde h_A=C_{A+}(\bm\alpha)h_+ + C_{A\times}(\bm\alpha)h_\times,
\quad A\in\{+,\times,b,l,x,y\},
\label{eq:linear_map}
\end{equation}
with $\bm\alpha=(\aone,\atwo,\athree)$. The coefficients depend on the rotation order because finite rotations in distinct planes do not commute. We denote $R_{yw}R_{xw}$ and $R_{xw}R_{yw}$ by Cases 1.1 and 1.2, respectively; these are two realizations of Case 1 in Ref.~\cite{Liu:2022cwb}. Here $\athree=0$, and only tensor and breathing components can be present. Cases 2.1--2.4 correspond, respectively, to $R_{yw}R_{zw}R_{xw}$, $R_{xw}R_{zw}R_{yw}$, $R_{zw}R_{yw}R_{xw}$, and $R_{zw}R_{xw}R_{yw}$, with the rightmost rotation applied first. These are realizations of Case 2 in Ref.~\cite{Liu:2022cwb}, and all six apparent polarizations can be present. The rotation matrices and explicit projected amplitudes for each ordering are given in Sec.~\ref{sec:sm_method} of the Appendix.

The detector itself is unchanged by this projection. For interferometer $I$, the measured strain is obtained by applying the standard antenna patterns to the projected waveforms:
\begin{equation}
\widetilde h_I(f;\bm\theta,\bm\alpha)=\sum_A F_I^A(\Omega,\psi)\widetilde h_A(f;\bm\theta,\bm\alpha),
\label{eq:detector_response}
\end{equation}
where $F_I^A$ is the antenna pattern for polarization $A$, $\Omega$ and $\psi$ specify the observed sky direction and polarization angle, respectively, and $\bm\theta$ contains the ordinary compact-binary parameters, including masses and spins. For a fixed observed direction and basis, $F_I^A=D_I^{ij}e^A_{ij}$ is determined by the unchanged detector tensor $D_I^{ij}$ and the basis tensor $e^A_{ij}$. The coefficients $C_{A+}$ and $C_{A\times}$ encode the dependence of the projected amplitudes on the extra-dimensional angles.

Substituting \cref{eq:linear_map} into \cref{eq:detector_response} gives an equivalent representation,
\begin{equation}
\begin{aligned}
\widetilde h_I &= \mathcal F_{I+}h_+ + \mathcal F_{I\times}h_\times,\\
\mathcal F_{IP}(\Omega,\psi,\bm\alpha)
&=\sum_A F_I^A(\Omega,\psi)C_{AP}(\bm\alpha),
\quad P\in\{+,\times\}.
\end{aligned}
\label{eq:effective_weights}
\end{equation}
The $\mathcal F_{IP}$ combine the geometric projection coefficients with the standard antenna patterns. For fixed rotation angles, each projected waveform evolves as a fixed linear combination of $h_+$ and $h_\times$.

At $\bm\alpha=\bm0$, the projected amplitudes satisfy
\begin{equation}
\begin{aligned}
\widetilde h_+&=h_+,\qquad \widetilde h_\times=h_\times,\\
\widetilde h_b&=\widetilde h_l=\widetilde h_x=\widetilde h_y=0.
\end{aligned}
\label{eq:gr_limit}
\end{equation}
Thus, the measured strain reduces to the standard four-dimensional GR prediction, $h_I^{\mathrm{GR}}=F_I^+h_+ + F_I^\times h_\times$. We refer to this as the four-dimensional GR waveform limit. Our analysis holds the GR waveform evolution fixed while varying the geometric projection angles.

\section{Data and Bayesian inference}\label{sec:methods}
We selected events from the first two publicly released parts of the fourth observing run by requiring a network signal-to-noise ratio (SNR) greater than 30 and coincident detection by at least the two LIGO detectors, H1 and L1. The five selected events are GW231226, GW240920, GW240925, GW241127, and GW250114. Their full identifiers, detector networks, and network SNRs are listed in Table~\ref{tab:event_sample} of the Appendix. Cases 1.1 and 1.2 were tested with all five events. We additionally tested Cases 2.1--2.4 with GW250114 and GW240920. GW250114 has the largest network SNR in the sample and gives the tightest constraints in the two-angle analysis.

We performed coherent Bayesian parameter estimation in the frequency domain. For detector data $d_I$ and stationary Gaussian noise,
\begin{equation}
\ln\mathcal L=-\frac12\sum_I\langle d_I-\widetilde h_I\mid d_I-\widetilde h_I\rangle_I+\mathrm{const.},
\label{eq:likelihood}
\end{equation}
where the noise-weighted inner product used the event-specific power spectral density (PSD). We used Bilby~2.8.0 \cite{Ashton:2018jfp} with Dynesty~3.0.0 \cite{Speagle:2019ivv}, 1000 live points, the acceptance-walk proposal with $\texttt{naccept}=60$, and 128 parallel CPU processes. No likelihood marginalization was applied. For every event, the tensor modes $h_+$ and $h_\times$ were generated with \texttt{IMRPhenomXPHM-SpinTaylor} \cite{Colleoni:2024knd}. 
We applied the geometric mixing in \cref{eq:linear_map} to these tensor waveforms and evaluated the strain using the unchanged antenna patterns in \cref{eq:detector_response}. All other settings followed the corresponding LIGO--Virgo--KAGRA (LVK) event analysis, including data conditioning, the analysis band, event-specific PSDs, calibration-uncertainty treatment, and priors on the ordinary binary parameters (including spins); these inputs were taken from the LVK posterior releases for these events. The lower frequency cutoff was 20~Hz.

We assigned independent uniform priors
\begin{equation}
\alpha_i\sim\mathcal U(-\pi/2,\pi/2),\qquad i=1,2,3.
\label{eq:angle_prior}
\end{equation}
The rotations are periodic over $2\pi$, and the angle maps contain multiple parameter combinations that reproduce the four-dimensional GR waveform. We use the local angular domain in \cref{eq:angle_prior}, centered on $\bm\alpha=\bm0$, for all reported constraints. For the tensor ansatz in \cref{eq:natural_tensor}, $\aone=\atwo=0$ leaves the projected polarization tensor unchanged for any $\athree$ in all four three-angle orderings.

\begin{table}[tbp]
\centering
\caption{Posterior medians and symmetric-tail 90\% credible intervals in radians for the two-angle analyses. The last column gives the joint GR quantile $\mathcal Q_{\mathrm{GR}}$, defined in \cref{eq:gr_quantile}; values below 90\% place the origin inside the estimated 90\% HPD region.}
\label{tab:alpha12}
\setlength{\tabcolsep}{4pt}
\begin{tabular}{@{}llccr@{}}\toprule
Case&Event&$\aone$&$\atwo$&\shortstack{$\mathcal Q_{\mathrm{GR}}$\\(\%)}\\\midrule
\multirow{5}{*}{1.1}
&GW231226&$0.01^{+0.86}_{-0.88}$&$-0.07^{+0.93}_{-0.84}$&0.34\\
&GW240920&$-0.07^{+0.79}_{-0.73}$&$-0.01^{+0.78}_{-0.83}$&3.69\\
&GW240925&$-0.02^{+0.77}_{-1.11}$&$0.06^{+1.17}_{-0.71}$&1.31\\
&GW241127&$-0.19^{+1.57}_{-0.84}$&$-0.30^{+1.60}_{-0.48}$&53.99\\
&GW250114&$0.00^{+0.57}_{-0.56}$&$0.01^{+0.51}_{-0.52}$&0.60\\
\midrule
\multirow{5}{*}{1.2}
&GW231226&$-0.09^{+0.91}_{-0.64}$&$-0.04^{+0.82}_{-0.88}$&2.26\\
&GW240920&$-0.07^{+1.00}_{-0.69}$&$0.07^{+0.81}_{-0.92}$&3.10\\
&GW240925&$0.14^{+0.69}_{-0.96}$&$0.01^{+0.71}_{-0.67}$&10.84\\
&GW241127&$0.23^{+0.63}_{-1.48}$&$0.29^{+0.71}_{-1.66}$&47.36\\
&GW250114&$-0.03^{+0.61}_{-0.53}$&$0.00^{+0.68}_{-0.73}$&0.55\\
\bottomrule
\end{tabular}
\end{table}

To quantify the location of the GR reference point within each angle posterior, we use a density-ranking GR quantile, following the contour-based construction of Ref.~\cite{LIGOScientific:2021sio}. For a normalized density $p$ on $\bm a=(\aone,\atwo)$, we define
\begin{equation}
\mathcal Q[p]=\Pr_{\bm a\sim p}\!\big[p(\bm a)>p(\bm0)\big].
\label{eq:gr_quantile}
\end{equation}
The quantile is the posterior probability contained in regions with density higher than that at the GR reference point. Smaller values therefore place the GR reference point closer to the high-density portion of the posterior, while the widths of the angle intervals quantify the constraint strength.

We report $\mathcal Q_{\mathrm{GR}}=\mathcal Q[p_2]$ for the two-angle joint posterior $p_2$, and $\mathcal Q^{\mathrm{marg}}_{\mathrm{GR}}=\mathcal Q[p_{\mathrm{marg}}]$ for the three-angle fit after integrating out $\athree$. Both statistics describe the GR reference $\aone=\atwo=0$ in a two-dimensional posterior; the latter integrates along the GR-equivalent line over the allowed values of $\athree$. For continuous densities with no finite-probability plateau at the reference density, a quantile below $0.9$ places the origin inside the corresponding 90\% highest-posterior-density (HPD) region. We estimate these quantiles from all equal-weight posterior draws using a full-covariance two-dimensional Gaussian kernel density estimate (KDE) with Scott's bandwidth, separately for each event and rotation ordering. The definitions, marginalization procedure, and numerical sensitivity checks are given in Sec.~\ref{sec:gr_quantiles} of the Appendix.

\begin{table}[tbp]
\centering
\caption{Posterior medians and symmetric-tail 90\% credible intervals (radians) for the three-angle analyses of GW250114 and GW240920 under Cases 2.1--2.4. The last column gives $\mathcal Q^{\mathrm{marg}}_{\mathrm{GR}}$ in percent for the two-dimensional $(\aone,\atwo)$ posterior after marginalizing over $\athree$. Event labels use the short names defined in Table~S1 of the Supplemental Material.}
\label{tab:alpha123}
\setlength{\tabcolsep}{2.5pt}
\begin{tabular}{@{}llcccr@{}}
\toprule
Case & Event & $\aone$ & $\atwo$ & $\athree$ & \shortstack{$\mathcal Q^{\mathrm{marg}}_{\mathrm{GR}}$\\(\%)}\\
\midrule
\multirow{2}{*}{2.1}
& GW250114 & $0.01^{+0.55}_{-0.49}$ & $0.05^{+0.67}_{-0.79}$ & $-0.02^{+1.09}_{-1.33}$&3.22\\
& GW240920 & $0.16^{+0.85}_{-0.89}$ & $-0.13^{+1.51}_{-1.11}$ & $0.19^{+1.27}_{-1.62}$&2.81\\
\midrule
\multirow{2}{*}{2.2}
& GW250114 & $-0.03^{+0.44}_{-0.48}$ & $-0.01^{+0.43}_{-0.53}$ & $-0.04^{+1.21}_{-1.24}$&0.61\\
& GW240920 & $0.21^{+0.98}_{-0.73}$ & $0.11^{+1.02}_{-0.82}$ & $0.18^{+1.18}_{-1.42}$&0.91\\
\midrule
\multirow{2}{*}{2.3}
& GW250114 & $0.00^{+0.47}_{-0.45}$ & $-0.05^{+0.64}_{-0.64}$ & $0.01^{+1.27}_{-1.28}$&2.83\\
& GW240920 & $0.08^{+0.62}_{-0.57}$ & $0.03^{+0.76}_{-0.76}$ & $0.11^{+1.22}_{-1.42}$&8.57\\
\midrule
\multirow{2}{*}{2.4}
& GW250114 & $0.00^{+0.55}_{-0.61}$ & $0.02^{+0.48}_{-0.53}$ & $0.02^{+1.38}_{-1.36}$&4.64\\
& GW240920 & $0.04^{+0.71}_{-0.72}$ & $-0.02^{+0.53}_{-0.62}$ & $-0.04^{+1.34}_{-1.30}$&0.38\\
\bottomrule
\end{tabular}
\end{table}

\section{Results}\label{sec:results}
\subsection{Propagation within the observed spatial subspace}
\begin{figure*}[t]
\centering
\includegraphics[width=0.48\textwidth]{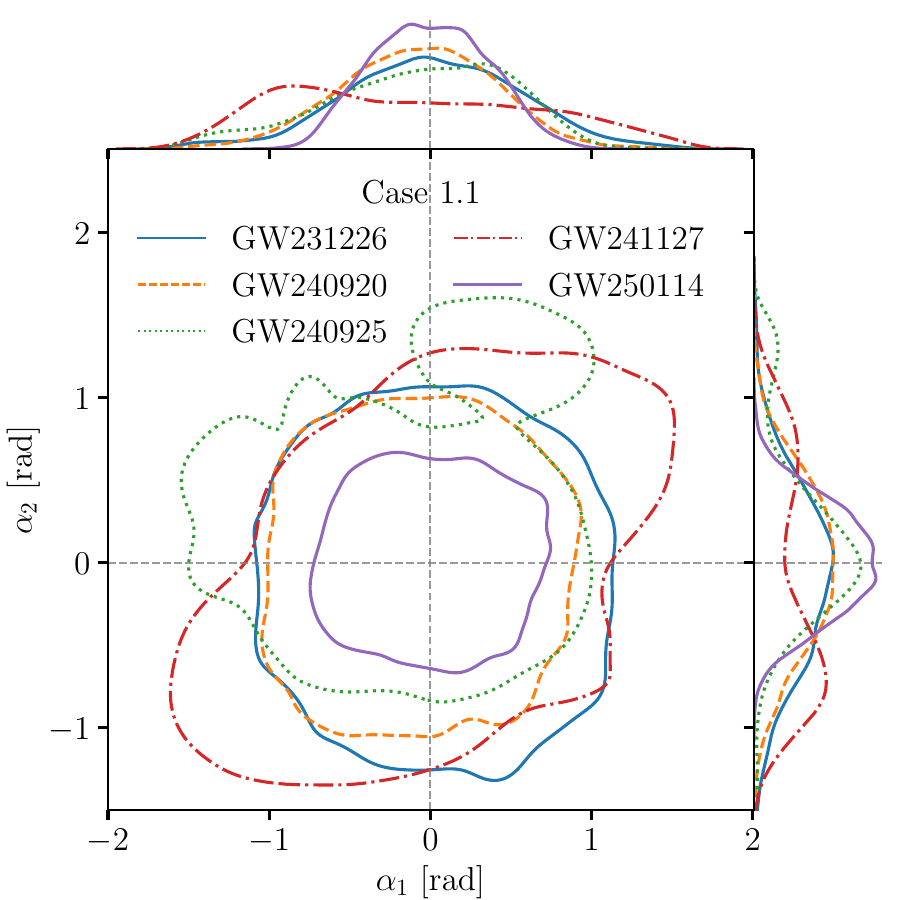}\hfill
\includegraphics[width=0.48\textwidth]{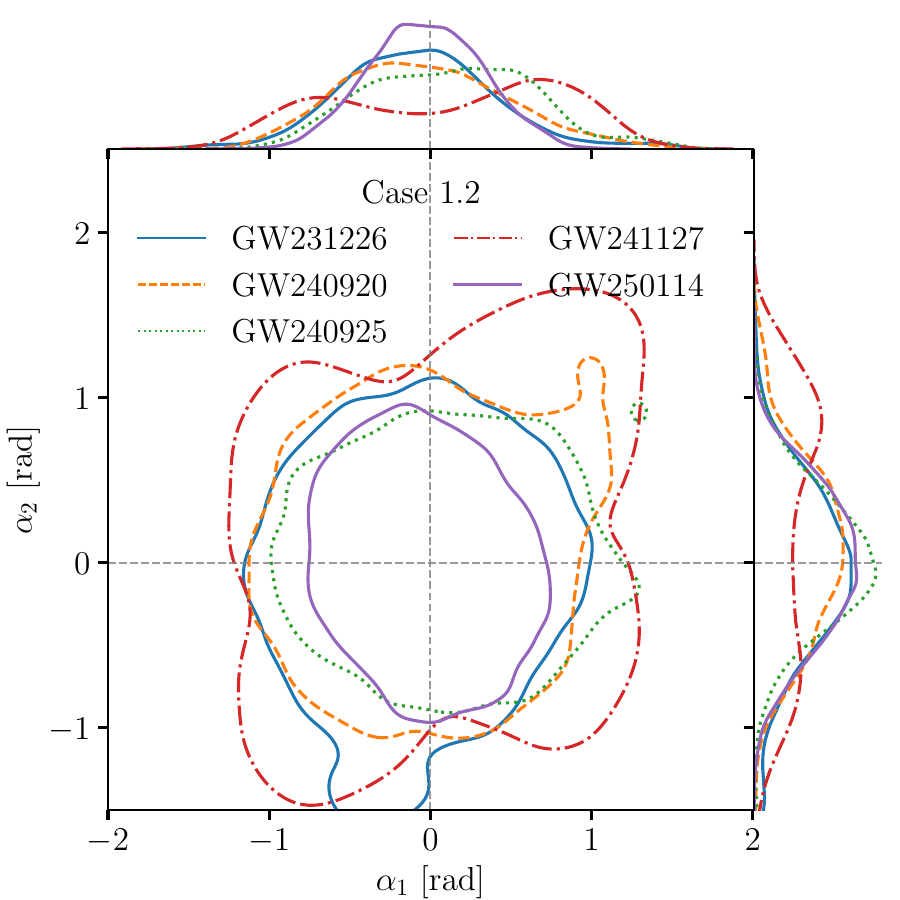}
\caption{Joint posteriors of $\aone$ and $\atwo$ for Cases 1.1 (left) and 1.2 (right). Curves enclose the 90\% credible regions; marginal distributions appear above and to the right. Dashed gray lines mark the four-dimensional GR waveform limit ($\alpha_1=\alpha_2=0$). Event labels use the short names defined in Table~\ref{tab:event_sample} of the Appendix.}
\label{fig:alpha12_joint}
\end{figure*}

The two-dimensional posteriors are shown in \cref{fig:alpha12_joint}, with one-dimensional intervals and joint GR quantiles in \cref{tab:alpha12}. The ten values of $\mathcal Q_{\mathrm{GR}}$ range from $0.34\%$ to $53.99\%$, placing the GR origin within the estimated 90\% HPD region for every event and both rotation orders. GW241127 has the largest values, $53.99\%$ and $47.36\%$ in Cases 1.1 and 1.2, respectively. The other eight quantiles lie between $0.34\%$ and $10.84\%$. The larger quantiles for GW241127 reflect its more pronounced off-origin and multimodal posterior structure, visible in the joint contours and marginal distributions in \cref{fig:alpha12_joint}. These off-origin concentrations of posterior probability leave a larger fraction of the posterior at densities above that of the GR origin, yielding larger $\mathcal Q_{\mathrm{GR}}$ values than for the other events.

GW250114, the highest-SNR event, gives the tightest Case 1.1 intervals, $\aone=0.00^{+0.57}_{-0.56}\,\mathrm{rad}$ and $\atwo=0.01^{+0.51}_{-0.52}\,\mathrm{rad}$. Its quantiles of $0.60\%$ and $0.55\%$ place the GR origin near the highest-density part of the two joint posteriors.

\subsection{Propagation tilted into the extra dimension}
\begin{figure*}[t]
\centering
\includegraphics[width=0.93\textwidth]{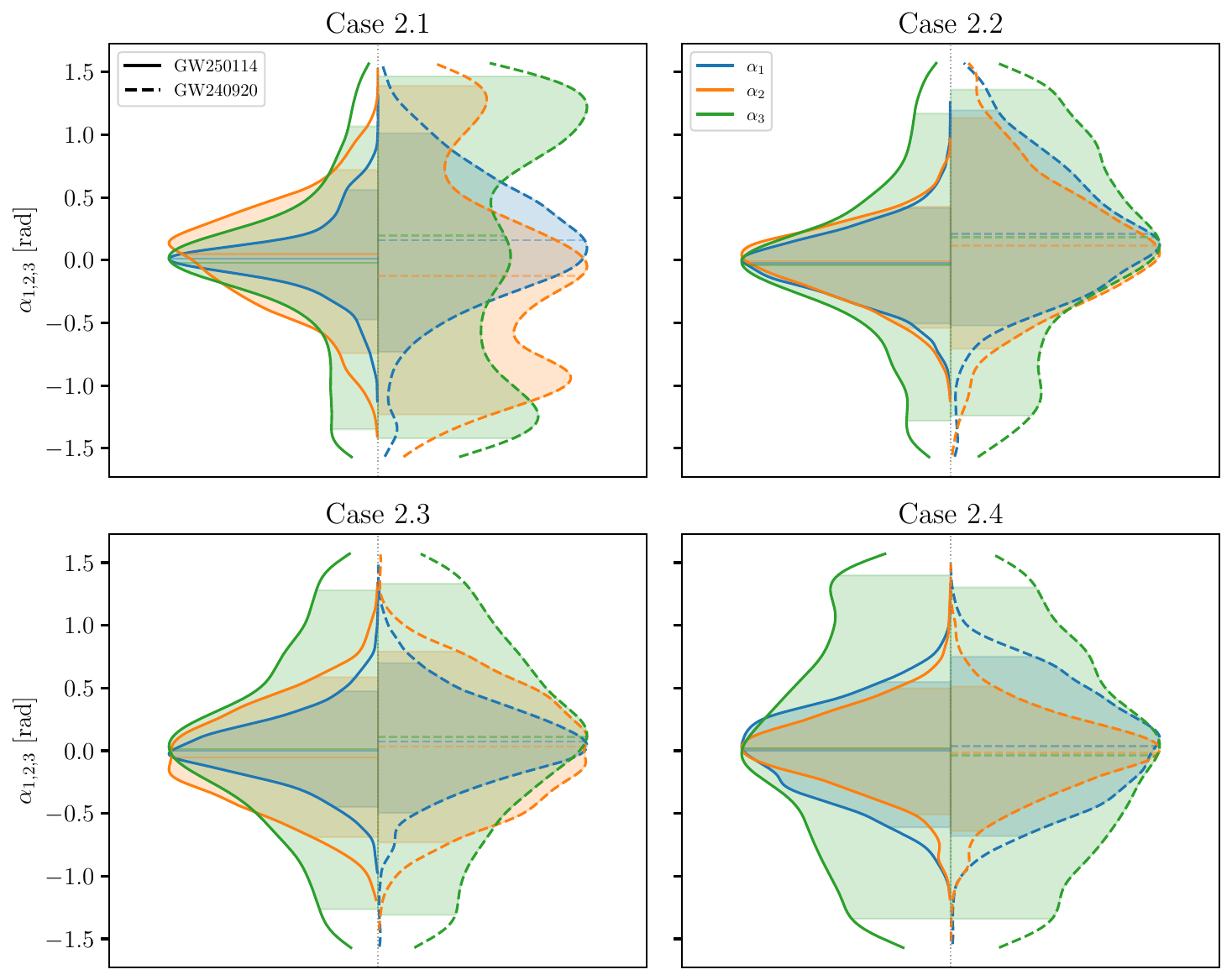}
\caption{Marginal posterior densities of $\aone$, $\atwo$, and $\athree$ for the four three-angle rotation orderings. Solid curves on the left show GW250114, and dashed curves on the right show GW240920. Blue, orange, and green denote $\aone$, $\atwo$, and $\athree$, respectively; horizontal colored markers indicate the posterior medians. Zero is the four-dimensional GR waveform limit.}
\label{fig:alpha123_violin}
\end{figure*}

Allowing the propagation direction to rotate into the extra dimension introduces vector and longitudinal polarization components and broadens the geometric degeneracies. The one-dimensional posteriors in \cref{fig:alpha123_violin} remain compatible with zero for GW250114 and GW240920 in all four rotation orderings. To assess the GR reference jointly in the transverse angles, \cref{tab:alpha123} also reports $\mathcal Q^{\mathrm{marg}}_{\mathrm{GR}}$ after marginalization over $\athree$. For GW250114, these values are $3.22\%$, $0.61\%$, $2.83\%$, and $4.64\%$ in Cases 2.1--2.4, respectively; for GW240920, they are $2.81\%$, $0.91\%$, $8.57\%$, and $0.38\%$. Thus, the origin lies well inside the estimated 90\% HPD region of all eight marginalized two-dimensional posteriors.

As shown by the intervals in \cref{tab:alpha123}, $\athree$ is generally less constrained than the transverse-plane angles, with intervals spanning much of the adopted local domain. Current data therefore constrain transverse-plane mixing more strongly than the propagation tilt.

\section{Discussion}\label{sec:discussion}
All 18 event--ordering analyses place the GR reference inside the estimated two-dimensional 90\% HPD region, consistent with the one-dimensional angle intervals and previous polarization tests. In the projection model, all apparent polarization components are fixed linear combinations of the same $h_+$ and $h_\times$ waveforms. The bounds therefore test the geometric relation between these components using the standard detector response.

Each of the six cases defines an angle-to-polarization map through its ordered product of noncommuting rotations. The corresponding posteriors describe the angle constraints for each ordering and its adopted prior. The exact degeneracy along $\aone=\atwo=0$, where the polarization waveform is independent of $\athree$, helps explain the broad propagation-tilt intervals.

The GR origin remains inside the estimated 90\% HPD region under all reported bandwidth and prior-boundary sensitivity checks (see Sec.~\ref{sec:gr_quantiles} of the Appendix). The largest boundary-reflection shift is $0.98$ percentage points, for GW241127 in Case 1.2, and the largest bandwidth-induced shift is $3.72$ percentage points. These changes characterize the numerical sensitivity of the density-ranking estimates.

The analysis uses two natural-frame tensor amplitudes, four-dimensional GR source waveforms, and the local angular prior in \cref{eq:angle_prior}. The inferred posteriors also depend on waveform accuracy, PSD estimates, calibration, and event selection. Extensions can explore the full periodic angular domain, additional higher-dimensional polarizations, and source and propagation models for specific extra-dimensional scenarios.


\section{Conclusion}\label{sec:conclusion}
We have developed and applied a test of apparent extra GW polarizations arising from the projection of a two-mode tensor wave in five-dimensional spacetime onto the observed three-dimensional spatial subspace. The projected components are fixed linear combinations of standard GR tensor waveforms and enter the measured strain through the standard detector antenna patterns. At zero rotation angles, the waveform recovers the four-dimensional GR prediction.

Frequency-domain Bayesian analyses of five compact-binary events constrain two rotation angles for two orderings that preserve propagation within the observed subspace. Analyses of GW250114 and GW240920 additionally constrain a propagation-tilt angle for four further orderings. The ten joint two-angle GR quantiles span $0.34\%$--$53.99\%$, while the eight quantiles obtained after marginalizing over the third angle span $0.38\%$--$8.57\%$. The GR reference lies within its corresponding estimated two-dimensional 90\% HPD region in every case, and this result is stable under the reported KDE sensitivity checks. The one-dimensional intervals constrain transverse-plane mixing more tightly than propagation tilt. Together, these results characterize the allowed geometric misalignment and are consistent with the GR waveform under the adopted model and angular prior.

By incorporating the correlations implied by projection, this approach provides a polarization-based test of extra-dimensional geometry complementary to propagation, amplitude-leakage, and black-hole-spectroscopy tests. Future analyses can combine larger event samples and more independent detector orientations with an explicit treatment of global angular degeneracies and waveform and calibration systematics. Source and propagation models for specific extra-dimensional scenarios can further connect these geometric constraints to the specific underlying theory.

This work made use of the following software packages: \texttt{astropy} \cite{astropy:2013, astropy:2018, astropy:2022}, \texttt{matplotlib} \cite{Hunter:2007}, \texttt{numpy} \cite{numpy}, \texttt{python} \cite{python}, \texttt{scipy} \cite{2020SciPy-NMeth, scipy_12522488}, \texttt{Bilby} \cite{Ashton:2018jfp, Bilby_2602178}, \texttt{PyCBC} \cite{Allen:2005fk}, \texttt{corner} \cite{corner-Foreman-Mackey-2016, corner.py_14209694}, \texttt{h5py} \cite{collette_python_hdf5_2014, h5py_7560547}, \texttt{pandas} \cite{mckinney-proc-scipy-2010, pandas_13819579}, \texttt{Jupyter} \cite{2007CSE.....9c..21P, kluyver2016jupyter}, \texttt{lalsuite} \cite{lalsuite}, \texttt{swiglal} \cite{swiglal}, and \texttt{GWpy} \cite{gwpy,Macleod:2021goi}.

\begin{acknowledgments}
This work was supported by the National Natural Science Foundation of China under grant numbers 12603066, 12405065, 12465013, 12573042, and 12475056.
H.-T. Wang is also supported by the Natural Science Foundation of Liaoning Province (Doctoral Scientific Research Start-up Fund, Grant No. 2025-BS-0065) and the Fundamental Research Funds for the Central Universities at Dalian University of Technology.
L.\ Shao was supported by the Max Planck Partner Group Program funded by the Max Planck Society.
This research made use of data and software obtained from the Gravitational Wave Open Science Center, a service of LIGO Laboratory, the LIGO Scientific Collaboration, the Virgo Collaboration and KAGRA.
\end{acknowledgments}
\AddToHook{env/thebibliography/begin}{\raggedright}
\bibliographystyle{apsrev4-2}
\bibliography{ref,softs}

\clearpage
\onecolumngrid
\appendix
\setcounter{secnumdepth}{2}

\section{Higher-dimensional projection}\label{sec:sm_method}
We consider a five-dimensional spacetime with four spatial coordinates $\{ x,y,z,w \}$, where $w$ denotes the additional spatial dimension.
To isolate the geometric projection effect, we adopt a natural-frame gravitational wave (GW) with two tensor amplitudes in the $x$--$y$ plane and propagation along the $z$ axis. Its polarization tensor is
\begin{align}
    p=\left(\begin{array}{cccc}
        h_+ & h_\times & 0 & 0   \\
        h_\times &  -h_+ & 0 & 0 \\
        0 & 0 & 0 & 0 \\
        0 & 0 & 0 & 0
    \end{array}\right) \ .
\end{align}
The relative orientation of the natural GW frame and the observer's three-dimensional spatial subspace determines the observable block of the higher-dimensional perturbation. We parametrize this relative orientation by rotations in the $x$--$w$, $y$--$w$, and $z$--$w$ planes, respectively:
\begin{align}
        R_{xw}=\left(\begin{array}{cccc}
        \cos\alpha_1 & 0  & 0 & -\sin\alpha_1  \\
         0 & 1 & 0 & 0  \\
         0 & 0 & 1 & 0 \\
         \sin\alpha_1 & 0 & 0 & \cos\alpha_1 
    \end{array} \right) ,  \\
        R_{yw}=\left(\begin{array}{cccc}
         1 & 0 & 0 & 0  \\
         0 & \cos\alpha_2  & 0 & -\sin\alpha_2  \\
         0 & 0 & 1 & 0 \\
         0 &  \sin\alpha_2  & 0  & \cos\alpha_2 
    \end{array} \right) , \\
        R_{zw}=\left(\begin{array}{cccc}
         1 & 0 & 0 & 0  \\
         0 & 1 & 0 & 0 \\
         0 & 0 & \cos\alpha_3   & -\sin\alpha_3  \\
         0 & 0 &  \sin\alpha_3    & \cos\alpha_3 
    \end{array} \right)  .
\end{align}
Here, $\alpha_1$ and $\alpha_2$ describe the mixing of the two transverse directions with the extra dimension, whereas $\alpha_3$ describes the mixing of the propagation direction with the extra dimension.

For a specified sequence of rotations, the polarization in the observer frame is $\widetilde p=R p R^{\mathsf T}$, where $R$ is the ordered product of the relevant rotation matrices, with the rightmost factor acting first. 
Since rotations in different planes generally do not commute, changing their order can lead to different projected polarization amplitudes.
We assume that the detector is confined to the observed spatial subspace and retains its standard coupling to the wave. It therefore samples only the $3\times3$ block associated with the observable coordinates.
We decompose this block as
\begin{align}
    {}^{(3)}\widetilde p = \left(\begin{array}{ccc}
        \tilde{h}_+ +\tilde{h}_b & \tilde{h}_\times & \tilde{h}_x    \\
        \tilde{h}_\times &  \tilde{h}_b-\tilde{h}_+  & \tilde{h}_y  \\
        \tilde{h}_x & \tilde{h}_y & \tilde{h}_l 
    \end{array}\right) \ .
\end{align}
Here, tilded quantities refer to the spatial frame of a four-dimensional observer, and untilded quantities refer to the natural GW frame.

\subsection{Case 1: propagation within the observable subspace}

We first consider the case in which the GW propagation direction remains along the observable $z$ axis, which corresponds to Case 1 in Ref.~\cite{Liu:2022cwb}. 
Equivalently, no rotation in the $z-w$ plane is introduced, so that $\alpha_3=0$.

For Case 1.1, we first apply $R_{xw}$ and then $R_{yw}$, so $R=R_{yw}R_{xw}$. The resulting tensor and breathing modes are given by 
\begin{align}
    \tilde{h}_+ &= \frac{1}{2}( \cos^2 \alpha_1 +\cos^2 \alpha_2 
    -\sin^2 \alpha_1 \sin^2 \alpha_2 )h_+ 
     +\frac12 \sin\alpha_1 \sin 2\alpha_2 \ h_\times \ , 
     \\
    \tilde{h}_\times &= -\frac12 \sin2\alpha_1 \sin \alpha_2 \ h_+ +\cos\alpha_1 \cos \alpha_2 \ h_\times \ ,
    \\
    \tilde{h}_b &= \frac{1}{2}( \cos^2 \alpha_1 -\cos^2 \alpha_2 
    +\sin^2 \alpha_1 \sin^2 \alpha_2 )h_+ 
     -\frac12 \sin\alpha_1 \sin 2\alpha_2 \ h_\times \ .
\end{align}

For Case 1.2, the order is reversed, so $R=R_{xw}R_{yw}$, and the amplitudes are
\begin{align}
    \tilde{h}_+ &= \frac{1}{2}( \cos^2 \alpha_1 +\cos^2 \alpha_2 
    -\sin^2 \alpha_1 \sin^2 \alpha_2 )h_+ 
     -\frac12 \sin2\alpha_1 \sin \alpha_2 \ h_\times  \ ,
     \\
    \tilde{h}_\times &= \frac12 \sin\alpha_1 \sin2\alpha_2 \ h_+ +\cos\alpha_1 \cos \alpha_2 \ h_\times  \ ,
    \\
    \tilde{h}_b &= \frac{1}{2}( \cos^2 \alpha_1 -\cos^2 \alpha_2 
    -\sin^2 \alpha_1 \sin^2 \alpha_2 )h_+ 
     -\frac12 \sin2\alpha_1 \sin \alpha_2 \ h_\times \ .
\end{align}

Because the propagation direction is not mixed with $w$ in Case 1, the vector and longitudinal components vanish. 
Nevertheless, rotations of the transverse directions into the extra dimension modify the observable $x-y$ block and can generate an apparent breathing component.

\subsection{Case 2: propagation partly through the extra dimension}

We next allow the GW propagation direction to be mixed with the extra dimension by including a rotation $R_{zw}$ with nonzero $\alpha_3$. 
In this configuration, the projection can generate all six observable polarization patterns.

We consider the following four rotation orderings:

Case 2.1: $R_{xw}$, followed by $R_{zw}$ and then $R_{yw}$, so $R=R_{yw}R_{zw}R_{xw}$.

The projected amplitudes are given by 
\begin{align}
    \tilde{h}_+ &= \frac12 (\cos^2 \alpha_1 +\cos^2 \alpha_2 -\sin^2 \alpha_1 \sin^2 \alpha_2 \cos^2 \alpha_3 ) h_+ 
    +\frac12 \sin\alpha_1 \sin 2\alpha_2 \cos \alpha_3 \ h_\times \ ,
    \\
    \tilde{h}_\times &= -\frac12 \sin 2\alpha_1 \sin \alpha_2 \cos \alpha_3 \ h_+ +\cos\alpha_1 \cos \alpha_2 \ h_\times \ ,
    \\
    \tilde{h}_b &= \frac{1}{2} (  \cos^2 \alpha_1 -\cos^2 \alpha_2 +\sin^2 \alpha_1 \sin^2 \alpha_2 \cos^2 \alpha_3  ) h_+ -\frac12 \sin \alpha_1 \sin 2 \alpha_2 \cos \alpha_3 \  h_\times \ ,
    \\
    \tilde{h}_l &= \sin^2 \alpha_1 \sin^2 \alpha_3 \  h_+ \ ,
    \\
    \tilde{h}_x &= -\frac12 \sin 2 \alpha_1 \sin \alpha_3 \ h_+ \ ,
    \\
    \tilde{h}_y &= \frac12 \sin^2 \alpha_1 \sin\alpha_2 \sin 2 \alpha_3 \ h_+ -\sin\alpha_1 \cos \alpha_2 \sin\alpha_3 \ h_\times \ .
\end{align}

Case 2.2: $R_{yw}$, followed by $R_{zw}$ and then $R_{xw}$, so $R=R_{xw}R_{zw}R_{yw}$.

The projected amplitudes are given by 
\begin{align}
    \tilde{h}_+ =& \frac12 ( \cos^2 \alpha_1 +\cos^2 \alpha_2 -\sin^2\alpha_1 \sin^2 \alpha_2 \cos^2 \alpha_3  )h_+ 
    - \frac12 \sin 2\alpha_1 \sin \alpha_2 \cos \alpha_3 \ h_\times \ ,
    \\
    \tilde{h}_\times =& \frac12 \sin \alpha_1 \sin 2\alpha_2 \cos \alpha_3 \ h_+ +\cos\alpha_1 \cos \alpha_2 \ h_\times \ ,
    \\
    \tilde{h}_b = & \frac12 ( \cos^2 \alpha_1 -\cos^2 \alpha_2 -\sin^2\alpha_1 \sin^2 \alpha_2 \cos^2 \alpha_3  )h_+ 
    - \frac12 \sin 2\alpha_1 \sin \alpha_2 \cos \alpha_3 \ h_\times \ ,
    \\
    \tilde{h}_l =& -\sin^2 \alpha_2 \sin^2 \alpha_3 \ h_+ \ ,
    \\
    \tilde{h}_x =& -\frac12 \sin\alpha_1 \sin^2\alpha_2 \sin 2\alpha_3 \ h_+ - \cos\alpha_1 \sin\alpha_2 \sin\alpha_3 \ h_\times \ ,
    \\
    \tilde{h}_y = & \frac12 \sin 2\alpha_2 \sin \alpha_3 \ h_+ \ .
\end{align}

Case 2.3: $R_{xw}$, followed by $R_{yw}$ and then $R_{zw}$, so $R=R_{zw}R_{yw}R_{xw}$.

The projected amplitudes are given by 
\begin{align}
    \tilde{h}_+ = & \frac12 ( \cos^2\alpha_1 +\cos^2 \alpha_2 -\sin^2 \alpha_1 \sin^2 \alpha_2 ) h_+ 
    +\frac12 \sin\alpha_1 \sin 2\alpha_2 \ h_\times \ ,
    \\
    \tilde{h}_\times =& -\frac12 \sin2\alpha_1 \sin \alpha_2 \ h_+ + \cos\alpha_1 \cos \alpha_2 \ h_\times \ ,
    \\
    \tilde{h}_b = & \frac12 ( \cos^2\alpha_1 -\cos^2 \alpha_2 +\sin^2 \alpha_1 \sin^2 \alpha_2 ) h_+ 
    -\frac12 \sin\alpha_1 \sin 2\alpha_2 \ h_\times \ ,
    \\
    \tilde{h}_l =&  (\sin^2 \alpha_1 \cos^2 \alpha_2 -\sin^2 \alpha_2 )\sin^2 \alpha_3 \  h_+
    +\sin\alpha_1 \sin 2\alpha_2 \sin^2 \alpha_3 \ h_\times \ ,
    \\
    \tilde{h}_x =& -\frac12 \sin2\alpha_1 \cos\alpha_2 \sin\alpha_3 \ h_+ - \cos\alpha_1 \sin\alpha_2 \sin\alpha_3 \ h_\times \ ,
    \\
    \tilde{h}_y = & \frac12 (1+\sin^2 \alpha_1 )\sin2\alpha_2 \sin\alpha_3 \ h_+  -\sin\alpha_1 \cos 2\alpha_2 \sin \alpha_3 \ h_\times \ .
\end{align}

Case 2.4: $R_{yw}$, followed by $R_{xw}$ and then $R_{zw}$, so $R=R_{zw}R_{xw}R_{yw}$.

The projected amplitudes are given by 
\begin{align}
    \tilde{h}_+ = & \frac12 ( \cos^2 \alpha_1 +\cos^2 \alpha_2 -\sin^2 \alpha_1 \sin^2 \alpha_2  )h_+
    -\frac12 \sin 2\alpha_1 \sin\alpha_2 \ h_\times \ ,
    \\
    \tilde{h}_\times = & \frac12 \sin\alpha_1 \sin2\alpha_2 \ h_+ +\cos\alpha_1 \cos\alpha_2 \ h_\times \ ,
    \\
    \tilde{h}_b = & \frac12( \cos^2 \alpha_1 -\cos^2 \alpha_2 -\sin^2 \alpha_1 \sin^2 \alpha_2 ) h_+ 
    -\frac12 \sin 2\alpha_1 \sin\alpha_2 \ h_\times \ ,
    \\
    \tilde{h}_l = & (\sin^2\alpha_1 -\cos^2\alpha_1 \sin^2 \alpha_2) \sin^2\alpha_3 \ h_+ 
    + \sin2\alpha_1 \sin\alpha_2 \sin^2\alpha_3 \ h_\times \,
    \\
    \tilde{h}_x = & -\frac12 \sin2\alpha_1 (1+\sin^2 \alpha_2)\sin\alpha_3 \ h_+ 
    -\cos2\alpha_1 \sin\alpha_2 \sin\alpha_3 \ h_\times \ ,
    \\
    \tilde{h}_y = & \frac12 \cos\alpha_1 \sin 2\alpha_2 \sin \alpha_3 \ h_+ -\sin\alpha_1\cos\alpha_2 \sin\alpha_3 \ h_\times \ .
\end{align}

When $\alpha_3=0$, Cases 2.1 and 2.3 reduce to Case 1.1, while Cases 2.2 and 2.4 reduce to Case 1.2.

\subsection{Correlated polarizations and the unchanged detector model}

For every rotation sequence above, each apparent polarization in the observer frame is a linear combination of the two retained tensor amplitudes:
\begin{equation}
\widetilde h_A=C_{A+}(\bm\alpha)h_+ + C_{A\times}(\bm\alpha)h_\times,
\quad A\in\{+,\times,b,l,x,y\},
\label{eq:linear_map}
\end{equation}
where $\bm\alpha=(\alpha_1,\alpha_2,\alpha_3)$. The coefficients are fixed by the rotation sequence and angles, and the evolution of all six projected components is determined by $h_+$ and $h_\times$. We assume that these two amplitudes follow standard four-dimensional general relativity (GR) compact-binary templates, including spin effects.

When all rotation angles vanish, the natural and observer frames align and
\begin{equation}
\widetilde h_+=h_+,\qquad \widetilde h_\times=h_\times,\qquad
\widetilde h_b=\widetilde h_l=\widetilde h_x=\widetilde h_y=0.
\label{eq:gr_limit}
\end{equation}
This is the four-dimensional GR waveform limit, with the binary spin parameters retained. More generally, $R_{zw}$ acts on the vanishing $z$--$w$ block of $p$, so $R_{zw}pR_{zw}^{\mathsf T}=p$. Thus $\alpha_1=\alpha_2=0$ reproduces the GR polarization waveform for arbitrary $\alpha_3$ in all four three-angle cases.

For interferometer $I$, the measured strain is
\begin{equation}
\widetilde h_I=\sum_A F_I^A(\Omega,\psi)\widetilde h_A,
\qquad F_I^A=D_I^{ij}e^A_{ij},
\end{equation}
where $D_I^{ij}$ is the standard detector tensor and $e^A_{ij}$ is the polarization basis tensor for the specified observed sky direction and basis. The standard detector tensor and antenna patterns act on the projected incident amplitudes $\widetilde h_A$ to determine the measured strain.

One may equivalently write
\begin{equation}
\widetilde h_I=\mathcal F_{I+}h_+ + \mathcal F_{I\times}h_\times,
\qquad
\mathcal F_{IP}=\sum_A F_I^A C_{AP},\quad P\in\{+,\times\}.
\end{equation}
These effective weights combine the wave projection with the standard detector response. At zero angles they reduce to $F_I^+$ and $F_I^\times$, respectively, yielding the usual four-dimensional GR strain.

\section{Event sample}\label{sec:event_sample}

The analysis uses five compact-binary events from the first two publicly released parts of the fourth observing run, selected by requiring a network signal-to-noise ratio (SNR) greater than 30 and coincident observations by both LIGO detectors. Table~\ref{tab:event_sample} lists the full event identifiers, short names used throughout the paper, detector networks, and network SNRs. All five events enter the two-angle analyses in Cases 1.1 and 1.2. GW250114 and GW240920 also enter the three-angle analyses in Cases 2.1--2.4.

\begin{table}[htbp]
\centering
\caption{Events selected from the first two publicly released parts of the fourth observing run. All have network $\mathrm{SNR}>30$ and were observed by at least H1 and L1. H, L, and V denote LIGO Hanford (H1), LIGO Livingston (L1), and Virgo (V1), respectively.}
\label{tab:event_sample}
\begin{tabular}{lccc}
\toprule
Full event name & Short name & Detectors & Network SNR\\
\midrule
GW231226\_101520 & GW231226 & HL  & 34.7\\
GW240920\_124024 & GW240920 & HL  & 37.4\\
GW240925\_005809 & GW240925 & HLV & 31.2\\
GW241127\_061008 & GW241127 & HLV & 31.3\\
GW250114\_082203 & GW250114 & HL  & 78.6\\
\bottomrule
\end{tabular}
\end{table}

\section{GR quantiles and numerical evaluation}\label{sec:gr_quantiles}

\subsection{Posterior density ranking}

For each event and rotation ordering, we locate the GR reference $\bm a=\bm0$, where $\bm a=(\alpha_1,\alpha_2)$, within the corresponding two-dimensional posterior. The two-angle model fixes $\alpha_3=0$, while the three-angle model integrates over $\alpha_3$. The latter follows the GR-equivalent line $\alpha_1=\alpha_2=0$ identified in Sec.~I, along which the polarization waveform is independent of $\alpha_3$.

Let $p(\bm a)$ be a normalized density on $\mathcal D_2=[-\pi/2,\pi/2]^2$, conditional on the event data, rotation ordering, model, and prior. All other parameters have been marginalized. Following the density-contour construction used in the inspiral--merger--ringdown consistency test of Ref.~\cite{LIGOScientific:2021sio}, define
\begin{equation}
\mathcal Q[p]=\int_{\mathcal D_2}p(\bm a)\,
\mathbb{I}\!\left[p(\bm a)>p(\bm0)\right]\,d^2\bm a,
\label{eq:qgr_density_rank}
\end{equation}
where $\mathbb{I}[\cdot]$ is the indicator function, equal to $1$ if the enclosed condition holds and $0$ otherwise. For the joint two-angle posterior $p_2$, the reported statistic is $\mathcal Q_{\mathrm{GR}}=\mathcal Q[p_2]$. The quantile is the posterior probability contained in the region with density above the GR reference density.

For a continuous posterior with no finite-probability plateau at $p(\bm0)$, $\mathcal Q[p]$ is the probability enclosed by the equal-density contour through the origin, including any disconnected higher-density regions. The origin enters the highest-posterior-density (HPD) region at this credible content, so $\mathcal Q[p]<0.9$ places it inside the 90\% HPD region. The joint density retains correlations between the angles and accommodates non-Gaussian and multimodal posteriors.

\subsection{Marginalization in the three-angle model}

Let $p_3(\alpha_1,\alpha_2,\alpha_3)$ be the joint three-angle posterior after marginalization over the binary and other nuisance parameters. We first construct
\begin{equation}
p_{\mathrm{marg}}(\bm a)=\int_{-\pi/2}^{\pi/2}
p_3(\alpha_1,\alpha_2,\alpha_3)\,d\alpha_3,
\label{eq:qgr_marginal_density}
\end{equation}
and then evaluate
\begin{equation}
\mathcal Q^{\mathrm{marg}}_{\mathrm{GR}}
=\int_{\mathcal D_2}p_{\mathrm{marg}}(\bm a)\,
\mathbb{I}\!\left[p_{\mathrm{marg}}(\bm a)>p_{\mathrm{marg}}(\bm0)\right]\,d^2\bm a.
\label{eq:qgr_marginal_rank}
\end{equation}
Equation~(\ref{eq:qgr_marginal_rank}) ranks the two-dimensional density obtained after marginalizing over $\alpha_3$. Its HPD regions describe the location of the GR reference in the $(\alpha_1,\alpha_2)$ plane.

The inputs are equal-weight Bilby posterior samples from the dynesty analysis. Retaining the $(\alpha_1,\alpha_2)$ coordinates of every joint posterior row yields samples of $p_{\mathrm{marg}}$, with all sampled values of $\alpha_3$ contributing to the integral. The density $p_2$ describes the fit with $\alpha_3$ fixed, whereas $p_{\mathrm{marg}}$ describes the fit with $\alpha_3$ free and subsequently integrated out.

\subsection{Sample-based density estimation}

For either model, let $\{\bm a_i\}_{i=1}^{N}$ be all available equal-weight posterior draws projected onto the two-angle plane. We use a full-covariance Gaussian kernel density estimate (KDE),
\begin{equation}
\widehat p(\bm a)=\frac{1}{N}\sum_{i=1}^{N}
\frac{\exp\!\left[-\tfrac12(\bm a-\bm a_i)^{\mathsf T}H^{-1}(\bm a-\bm a_i)\right]}
{2\pi|H|^{1/2}},\qquad H=h^2S,\qquad h=N^{-1/6},
\label{eq:qgr_kde}
\end{equation}
where $S$ is the sample covariance matrix and $h$ is Scott's bandwidth factor in two dimensions. The density-estimation dimension is two for both fitted models. The reported estimator is
\begin{equation}
\widehat{\mathcal Q}=\frac{1}{N}\sum_{i=1}^{N}
\mathbb{I}\!\left[\widehat p(\bm a_i)>\widehat p(\bm0)\right].
\label{eq:qgr_estimator}
\end{equation}
The posterior draws supply the probability mass and the KDE supplies the density ordering. We use the same rows to fit the KDE and evaluate the empirical fraction in Eq.~(\ref{eq:qgr_estimator}), separately for each event and rotation ordering.
Tables I and II of the main text report $100\widehat{\mathcal Q}$, using the unreflected KDE at the nominal Scott bandwidth. 

\subsection{Boundary and bandwidth sensitivity}

The angular priors are uniform on $[-\pi/2,\pi/2]$. The main estimates use the unreflected Gaussian KDE. Since Gaussian kernels extend beyond the finite prior domain, we assess sensitivity to the boundaries using reflected evaluations. A common density normalization cancels in the ranking.

To assess boundary sensitivity, define $L=-\pi/2$, $U=\pi/2$ and the maps $T_0(a)=a$, $T_1(a)=2L-a$, and $T_2(a)=2U-a$. The diagnostic density is
\begin{equation}
\widehat p_{\mathrm{refl}}(\bm a)
=\sum_{(s_1,s_2)\in\{0,1,2\}^2}
\widehat p\!\left(T_{s_1}(a_1),T_{s_2}(a_2)\right),
\qquad \bm a\in\mathcal D_2.
\label{eq:qgr_boundary}
\end{equation}
This nine-term image sum approximates the boundary correction using the original two-dimensional KDE bandwidth. Both endpoints are treated as finite prior boundaries, and the density ranks are evaluated on the same posterior draws. Separate estimates with bandwidth factors $0.8h$, $h$, and $1.2h$ assess smoothing sensitivity.

Across the 18 reported analyses, boundary-reflection corrections change the quantiles by at most $0.98$ percentage points, while varying the KDE bandwidth between $0.8h$ and $1.2h$ changes them by at most $3.72$ percentage points. The largest bandwidth sensitivity occurs for GW240925 in Case 1.2, for which the quantile varies from $9.11\%$ to $14.56\%$. These variations do not affect the conclusion that the GR origin lies inside the estimated 90\% HPD region in any analysis.

\end{document}